# Toward Human-Centered Simulation Modeling for Critical Infrastructure Disaster Recovery Planning

Abbas Ganji
Department of Civil & Environmental Engineering
University of Washington
Seattle, WA, USA
ganjia@uw.edu

Scott Miles
Department of Human-Centered Design & Engineering
University of Washington
Seattle, WA, USA
milessb@uw.edu

*Abstract*—Critical infrastructure is vulnerable to a broad range of hazards. Timely and effective recovery of critical infrastructure after extreme events is crucial. However, critical infrastructure disaster recovery planning is complicated and involves both domain- and user-centered characteristics and complexities. Recovery planning currently uses few quantitative computer-based tools and instead largely relies on expert judgment. Simulation modeling can simplify domain-centered complexities but not the human factors. Conversely, human-centered design places end-users at the center of design. We discuss the benefits of combining simulation modeling with human-centered design and refer it as human-centered simulation modeling. Human-centered simulation modeling has the capability to make recovery planning simpler and more understandable for critical infrastructure and emergency management experts and other recovery planning decision-makers.

We qualitatively analyzed several resilience planning initiatives, post-disaster recovery assessments, and relevant journal articles to understand experts and decision-makers' perspectives. We propose a conceptual design framework for creating human-centered simulation models for critical infrastructure disaster recovery planning. This framework consists of three constructs: 1) user interaction with design features that end-users interact with, including model parameters assignment, decision-making support, task queries, and usability; 2) system representation that refers to system components, system interactions, and system state variables; and 3) computation core that represents computational methods required to perform processes.

*Keywords—resilience and recovery planning, critical infrastructure recovery, simulation modeling, human-centered design*

## I. INTRODUCTION

Critical infrastructure is vital to the functioning of communities; however, it is also vulnerable to a broad range of hazards. Critical infrastructure is required to stay functional, mitigate hazard impacts, or be minimally damaged during and in the aftermath of disasters [1, 2, and 3]. In practice, however, disasters often damage various components of critical infrastructure that are outdated and poorly maintained. Timely and effective recovery of critical infrastructure after extreme events is crucial. Recovery "involves the actions taken in the long term after the immediate impact of the disaster has passed to stabilize the community and to restore some semblance of normalcy" [4]. Major disruption of a sector of critical infrastructure and its recovery timeframe may impact the performance and recovery of other sectors greatly. Effective recovery entails understanding various aspects of the critical infrastructure disaster recovery process, such as vulnerability, recovery management, and recovery timeframe of damaged components.

The complexity of the recovery process makes it difficult for decision-makers to clearly understand the process, highlighting the need for better tools to better understand the process. The Federal Emergency Management Agency (FEMA) emphasizes the importance of tools in emergency management, noting that "innovative models and tools" are one of three strategic needs to accomplish recovery planning [5]. A well-known example of such a tool is Hazus, which was created by FEMA as "a nationally applicable standardized methodology that contains models for estimating potential losses from earthquakes, floods, and hurricanes" [6]. There is a lack of such tools for pre- or post-event disaster recovery planning. Recovery planning currently uses few quantitative computer-based tools and instead largely relies on expert judgment.

In last ten years, several resilience planning initiatives in the U.S. have brought experts in critical infrastructure and emergency management together. The purpose of these initiatives was to provide recommendations for decision-makers to shorten the recovery process. This included collaboratively estimating target recovery timeframes and expected recovery timeframes of infrastructure systems subjected to potential hazard scenarios. Although the initiatives were successful in gathering many experts from different disciplines to undertake collaborative resilience planning and offer extensive recommendations, analytical computer-based tools were rarely used to facilitate the planning process or associated decision-making [7-12]. This is not surprising given the lack of such analytical computer-based tools that have the potential to aid experts and decision-makers in collaborative recovery planning.

In this paper, we aim to propose a conceptual design framework for development and design of analytical human-centered simulation modeling to aid in critical infrastructure recovery planning. The framework is based on the characteristics of critical infrastructure disaster recovery planning, and the desires and limitations of experts and decision-makers. We first investigate characteristics of critical infrastructure disaster recovery planning. We then introduce human-centered simulation modeling as a paradigm for creating analytical computer-based tools for this purpose. We discuss why this approach has potential to effectively support and improve collaborative planning for critical infrastructure disaster recovery. Subsequently, we describe in detail the proposed conceptual design framework for development and design of human-centered simulation modeling for recovery planning. This framework is based on a qualitative analysis of literature on disaster recovery from end-user's perspective and can be used for creating simulation modeling for end-users to promote and facilitate recovery planning.

## II. CHARACTERISTICS OF CRITICAL INFRASTRUCTURE DISASTER RECOVERY PLANNING

Understanding critical infrastructure disaster recovery planning and shedding light on its complexity is essential. We discuss important characteristics of critical infrastructure recovery planning in three increasingly focused levels: 1) critical infrastructure in general and as a system, 2) critical infrastructure disaster recovery as a process, and 3) critical infrastructure disaster recovery planning.

As can be seen in Fig. 1, the characteristics of critical infrastructure disaster recovery planning can be located along a spectrum, with the two ends of domain- and user-centered aspects. The upper characteristics in Figure 1 are more domain-centered and the lower ones are more user-centered. Domain-centered aspect represents the characteristics of critical infrastructure and its recovery process that are less impacted by human factors. User-centered aspect, on the other end of the spectrum, focuses on the characteristics that are heavily influenced by human factors.

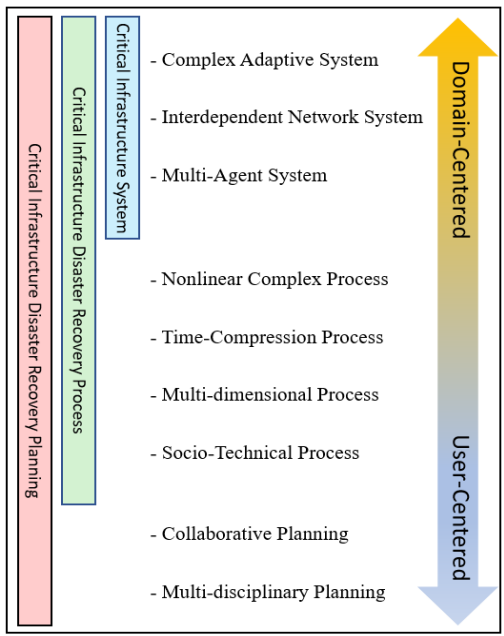

Fig. 1. Characteristics of critical infrastructure disaster recovery planning

### A. Critical Infrastructure as a System

Critical infrastructure systems are irreplaceable services and required capitals to offer those services [13, and 14]. Services are defined as "the link between capitals and their benefits to communities" [13]. The major characteristics of critical infrastructure systems are discussed below.

A.1. Complex Adaptive System: When a system is not complex, it is likely that an analyst would be able to predict the consequences of a change in the system. A large-scale system contains smaller systems within it, which are referred to as subsystems. As the number of subsystems and interrelationships increases, especially if the relationships in the system are nonlinear and adaptive, the system becomes too complicated to be easily understood. This is considered a complex system [15]. The consequences of changes in inputs are unpredictable in such cases. Critical infrastructure systems are complex adaptive systems since "they are all complex collections of interacting components in which change often occurs as a result of learning process" [16].

A.2. Interdependent Network System: There is remarkable evidence of infrastructure interdependency in the real world, cited in literature that demonstrates how a failure in one component causes several unanticipated degradations of other infrastructure sectors [17, and 18]. Critical infrastructure sectors are complex networks on their own, but even more complex when considering dependent and interdependent relationships with other sectors [16]. Rinaldy et al. (2001) divide interactions in critical infrastructure into dependency and interdependency. A dependent relationship is a unidirectional "linkage or connection between two infrastructures, through which the state of one infrastructure influences or is correlated to the state of the other." An interdependent relationship is bidirectional, meaning the state of each infrastructure correlates to the state of the other and the two depend on each other. Critical infrastructure interdependencies are categorized into four groups: physical, cyber, geographical, and logical [16].

A.3. Multi-Agent System: Critical infrastructure involves many agents from various organizations at different levels such as federal, state, county, and city, and is managed and controlled by public and private organizations. Critical infrastructure systems are interconnected, and individual agents' performance and decisions affect the entire network's performance. Agents act independently due to having potentially different goals and priorities. Thus, it is essential to take into account the multi-agent nature of critical infrastructure when planning for recovery to ensure that agents are able to cooperate to set and reach global goals with minimal supervision [19, and 20].

### B. Characteristics of the Critical Infrastructure Disaster Recovery Process

For the purposes of this paper, recovery is a long-term process that can be defined as returning to normal or reaching a better or new situation [3, and 21]. The main characteristics of critical infrastructure disaster recovery are provided below.

B.1. Nonlinear Dynamic Process: Disaster recovery is a nonlinear, unorderly, and dynamic process aimed at restoring the community to its normal pre-disaster conditions by reconstruction of damaged components. Recovery is a time-dependent process such that its timeframe does not change proportionally to input variables change [22].

B.2. Time-Compressed Process: In normal conditions, a low rate of loss of capital services is observed due to infrastructure components reaching the end of their life cycle and being replaced. However, unusually large and immediate loss of capital services occurring due to disasters results in an unusual increase in the rate of new capital services, with a corresponding increase in decisions, information flow, financing, and institutional formation. This increased pace of activity distinguishes the disaster recovery process from the normal process of replacing outdated capital services [23]. This situation "opens unusual opportunities for reorganizing or relocating capital facilities. Strategies of replacement may become available that would not be worth pursuing at normal rates of capital replacement" [23].

B.3. Multi-dimensional Process: Efforts have been made to explore, theorize, assess, and analyze recovery of sectors of critical infrastructure independently, such as built environment [24], business and economic [25, and 26], social [27, and 28], health care [29], transportation system [30], water and sewer system [31, and 32], and electric system [33]. However, critical infrastructure is heavily interdependent. The recovery process of a dimension of critical infrastructure impacts the recovery process of other dimensions. For example, recovery of the drinking water system depends on recovery of the power system because the former simply needs electricity to function. This makes critical infrastructure disaster recovery a multi-dimensional process [34, and 35]. Multi-dimensional recovery also causes different rates of recovery in different dimensions [36].

B.4. Socio-Technical Process: Critical infrastructure should be considered a socio-technical system [37], meaning that it has both a social and technical condition and there is a reciprocal relationship between its human and technical aspects such that "efficiency and humanity would not contradict each other" [38]. Critical infrastructure disaster recovery is influenced by these technical and social aspects. Government agencies, social communities, and politicians impact the recovery process, for example by allocating resources and determining priorities. This interaction between the social and technical is observed in the aftermath of disasters, both in the short-term (emergency response) and in the long-term phases of recovery [39]. Leavitt and Kiefer (2006), for instance, provide the human and political impacts on the critical infrastructure recovery failure that occurred after Hurricane Katrina due to decision-makers not understanding the technical complexity of infrastructure interdependency [39].

*C. Critical Infrastructure Disaster Recovery Planning*

Planning for critical infrastructure disaster recovery takes place by diverse groups of experts and stakeholders, who are not necessarily experts in all or any sectors of critical infrastructure. Planning for disaster recovery adds additional complexity that must be addressed when creating simulation modeling.

C.1. Collaborative Planning: Collaborative planning is engagement of government stakeholders, public and private business stakeholders, and community and organizational stakeholders in the process of planning. Collaborative planning can facilitate information sharing among stakeholders and the community, enhance decision-making, and raise the "community's ability to work toward collective goals" [40]. Experts in critical infrastructure and emergency management have undertaken several initiatives to envision resilience and recovery timeframes and provide recommendations to governmental decision-makers in the U.S. [40]. These initiatives have emphasized the highly collaborative nature of resilience planning. The National Institute of Standards and Technology (NIST) recommends forming a planning team as the first step toward community resilience planning for buildings and infrastructure systems [40].

C.2. Multi-disciplinary Planning: Because critical infrastructure systems and their recovery are multi-agent and multi-dimensional, recovery planning is required to be multi-disciplinary [41, and 42]. Multi-disciplinary planning gathers experts from diverse areas of expertise. These experts may have different technical languages and terminologies, priorities, and criteria. They also possibly have unequal levels of truthfulness and familiarity with analytical computer-based tools.

III. TOWARD HUMAN-CENTERED SIMULATION MODELING

The purpose of this paper is to propose a conceptual design framework for creating computer-based tools for critical infrastructure disaster recovery planning. These tools are required to be usable and understandable by critical infrastructure and emergency management experts who participate in recovery planning. In this paper, we refer to them as "end-users." To this end, we discussed the main characteristics of critical infrastructure disaster recovery planning, sorted along a spectrum with domain- and user-centered ends (Fig. 1). This list grounds the foundation of the conceptual design framework. It also indicates the complexities of recovery planning that require computer-based tools to address. Accordingly, we build our framework in two steps to consider both aspects.

*A. Simulation Modeling to Capture Domain-Centered Aspect*

Simulation modeling is capable of capturing the domain-centered characteristics of critical infrastructure recovery planning shown in Fig. 1. Simulation modeling of critical infrastructure explicitly represents the behavior or functioning of such networked and interdependent systems. It enables modelers to manipulate system details and explore the influence of different system characteristics. Simulation modeling is widely employed for modeling of critical infrastructure systems and their interdependencies [18 and 43]. It has been used to simulate different sectors of critical infrastructure disaster recovery such as power systems [44, and 45], water systems [46], and transportation networks [47]. Simulation modeling is also used for modeling interdependencies among infrastructure systems for modeling restoration in the aftermath of extreme events [48].

In general, the end-user's needs, expectations, and limitations are poorly addressed in the critical infrastructure recovery simulation modeling literature. Simulation models can be challenging for decision-makers, emergency managers, and critical infrastructure experts to comprehend [49]. End-users may have inadequate familiarity with and experience in using simulation modeling. Additionally, the ease of collaboration with other end-users affects how comfortable and willing they are to use simulation modeling.

*B. Human-Centered Design to Capture User-Centered Dimension*

The literature on socio-technical system design can help address this gap for designing simulation models for use in critical infrastructure recovery planning. Socio-technical system design is aimed at promoting, improving, and using the characteristics of socio-technical systems in system design, and has been developed in different ways over time [50]. Baxter and Sommerville (2011) provide seven categories of socio-technical system design approaches: 1) Soft system methodology, 2) Cognitive work analysis, 3) Socio-technical method for designing work systems, 4) Ethnographic workplace analysis, 5) Contextual design, 6) Cognitive systems engineering, and 7) Human-centered design. Baxter and Sommerville (2011) analyzed the seven approaches based on how well they cover three phases of the systems engineering life cycle—analysis, design, and evaluation—and a set of principles defined for their study [51]. They conclude that human-centered design is best suited for socio-technical system design.

Human-centered design is "a process of assuring that the concerns, values, and perceptions of all stakeholders in a design effort are considered and balanced" [52]. It facilitates innovative approaches for identifying and incorporating human (user) needs in the process of problem-solving. Norman (2013) states human-centered design is "the process of ensuring that people's needs are met, that the resulting product is understandable and usable, that it accomplishes the desired tasks, and that the experience of use is positive and enjoyable" [53]. Human-centered design is an iterative process that involves potential end-users throughout the development process. Functionally, human-centered design is often conducted in a process that repeats four overlapping steps until user needs are effectively met. These four steps are user research, prototyping, usability testing, and implementation.

While simulation modeling can capture domain-centered features to enable emergency management and critical infrastructure experts to undertake system and process monitoring and decision-making, human-centered design can be incorporated in the model development process to improve the usability of simulation models for end-users. We propose to combine human-centered design and simulation modeling and refer to this synthesis as human-centered simulation modeling.

IV. HUMAN-CENTERED SIMULATION MODELING DESIGN FRAMEWORK

In this section, we lay out a conceptual design framework for developing human-centered simulation models. We discussed the characteristics of recovery planning and the capability of human-centered simulation modeling for supporting critical infrastructure recovery planning in high level in previous sections. However, more information and details of end-users' concerns regarding recovery planning is needed to form and extend the conceptual design framework.

For this purpose, we collected and reviewed relevant data from three sources: 1) resilience planning initiatives, 2) post-disaster recovery assessments, and 3) research articles. We then qualitatively analyzed them to understand potential end-users' points of view related to the recovery and recovery planning. The collected data are briefly introduced below.

1. Resilience planning initiatives: Three initiatives have taken place in the U.S. in the last ten years to envision seismic community resilience on the state or city scales: the San Francisco Bay Area Planning and Urban Research Association (SPUR) Resilient City initiative, Resilient Washington State (RWS), and the Oregon Resilience Plan (ORP). These initiatives are valuable sources of information. They were performed by experts, managers, and decision-makers who would be the potential end-users of human-centered simulation models for critical infrastructure disaster recovery planning. Resilience planning is heavily connected to and has much in common with recovery planning, especially the pre-disaster recovery planning phase. Reviewing the initiatives provides insights into the objectives, concerns, and limitations of end-users [7-12].

The initiatives commonly aimed to establish and present the target recovery timeframe of various components of the community subjected to the expected seismic hazard, and estimate expected recovery timeframes of potentially damaged components. They also offer recommendations to improve community seismic resilience. The initiatives organized focus groups, for example a transportation group, and categorized the participants into these groups based on their areas of expertise. Each group or sector presented target and estimated expected recovery timeframes of services and components of the group. These estimates were obtained from debates, discussions, and participants' judgement. Noticeably, no analytical computer-based tools such as models were used in the resilience planning process. This shows the potential for using human-centered simulation modeling to support resilience and recovery planning [7-12].

2. Post-disaster recovery assessments: Another source of understanding real-world recovery processes and experts' perspectives is post-disaster recovery assessment reports published after investigation of post-disaster recovery of infrastructure disruptions. These after-action recovery assessments are usually prepared for governmental departments to assess efficiency of recovery processes and provide recommendations for infrastructure system operators to be prepared for future disasters. These documents are beneficial for our purposes because they assess practical planning operations performed by emergency managers and infrastructure system agents. These reports offer recommendations from various agents and organizational collaborations and identify the need for creating and using appropriate tools for damage assessment, recovery monitoring, and decision-making [54-59].

3. Research articles: Emergency management and infrastructure experts' experience is also addressed in research articles. Although we found representations of agents, practitioners, participants, and decision-makers to be poor in research studies, several articles do provide relevant information. We reviewed abstracts of papers published in and after 2000 in the journals *Natural Hazards Review* and *Earthquake Spectra* and identified articles that presented the experience and concerns of end-users. These studies investigate end-users via interviews and participatory studies, or by presenting frameworks for tools and usability testing [60-75].

We conducted qualitative content analysis of the literature described above to create a human-centered simulation modeling design framework shown in Fig. 2. Three main constructs emerged from the qualitative analysis are: user interaction, system representation, and computation core. Collectively, the three constructs include 11 elements, which are described below.

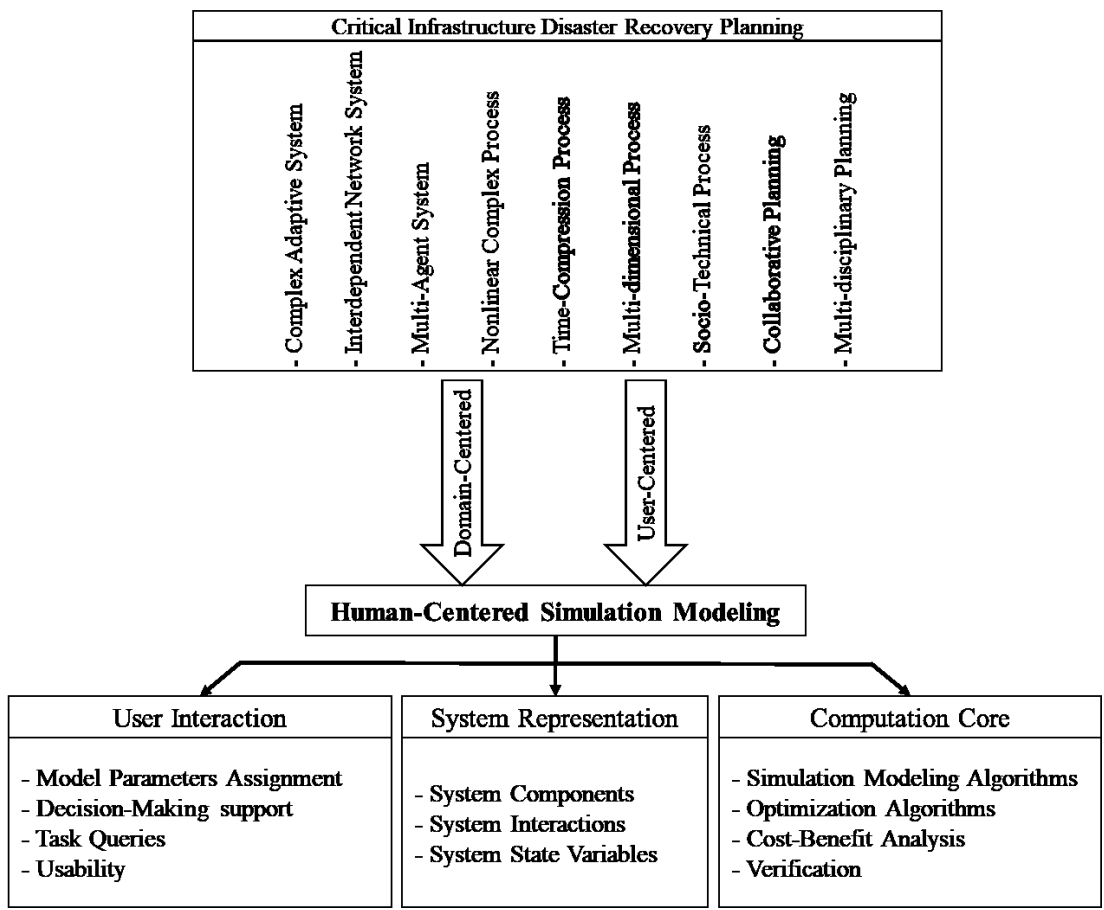

Fig. 2. Conceptual design framework of human-centered simulation modeling for critical infrastructure recovery planning

*A. User Interaction*

User interaction addresses design features that the end-user interacts with in human-centered simulation models. This construct has four elements, comprising model parameters assignment, decision-making support, task queries, and usability, as summarized in Table 1. The model parameters construct consists of three components: (a) hazard status parameters that provide hazard information such as scenario, size of disaster, and aspects of disaster (e.g., earthquake, liquefaction, landslide, hurricane, flood); (b) system status parameters such as vulnerability and resilience of components, damaged components and level of damage, time, cost, and resources required for recovery of damaged components, type of clients, and number of impacted clients; and (c) resource parameters that define system resourcefulness such as number of available crews, budget, materials, etc.

Decision-making support represents features that support the end-user's decision-making to prioritize and target goals built upon prioritization and target recovery timeframes. Prioritization represents features that enable the end-user to prioritize the recovery process. The end-user may plan to prioritize recovery of specific clients. For example, an end-user might desire to first recover critical buildings such as hospitals, or to prioritize damaged components whose recovery would provide services to more people. Also, target recovery timeframes of damaged services may be determined differently based on the end-user's decision-making.

Task queries point out information that the end-user desires to track within the recovery process and consists of time-variant indicators, critical path, and comparative analysis. Time-variant indicators enable the end-user to track desired recovery indicators over time such as recovery timeframes of components, sectors, or an entire modeled system; budget, cost, and resources over time; and social indicators. Critical path identifies critical recovery paths based on desired criteria such as finding the closest recovery path that provides services to a client or the least expensive path to recovery of selected clients. Comparative analyses facilitate end-user's comparison of the consequences of different parameters or decisions such as cost-benefit analysis, sensitivity analysis, and scenario analysis.

Usability represents the ease of use and learning by the end-user, including data navigation (e.g., simplicity of import and export of data with different formats, appropriate and understandable visualizations), help bar (e.g., memo, tutorial, item definition, guide documents), and knowledge transferability (i.e., transferability of organization and distribution of knowledge from researchers and tool developers to end-user and improvement of end-user's communication).

TABLE I. ELEMENT OF USER INTERACTION.

| Element | Component | Description and Example |
|---|---|---|
| Model Parameters Assignment | Hazard status parameters | Hazard scenario, size of disaster, secondary hazards |
| Model Parameters Assignment | System status parameters | Resilience of components, damaged components, clients |
| Model Parameters Assignment | Resources | Availability of crews, budget, materials |
| Decision-Making Support | Prioritization | Ability to prioritize recovery based on number of impacted people, type of clients, time, and cost |
| Decision-Making Support | Target assignment | Target timeframe recovery for components, sectors, and entire system |
| Task Queries | Time-Variant Indicators | Track recovery timeline, cost, budget, and resources timeframes, and social indicators |
| Task Queries | Critical path | Establishing critical path for recovery of specific components or clients |
| Task Queries | Comparative analyses | Cost-benefit, scenario-based, and sensitivity analyses |
| Usability | Data navigation | Simplicity of import and export of data in desired format, appropriate visualizations |
| Usability | Help bar | Memo, tutorial, item definition, documentation guide |
| Usability | Knowledge transfer | Transferability of organization and distribution of knowledge from researchers and tool developers to end-users, improvement of end-user's communications, information sharing among end-users, and using understandable terminology |

*B. System Representation*

"System" refers to entities, components, networks, and interconnections of a modeled infrastructure sector. Systems in this framework can be used for any type of infrastructure systems such as built or social systems. Systems can be conceptually broken down into system components, state variables, and interactions. System components represent (a) entities of systems under consideration such as electric power entities, water system entities, clients, and geographical information of entities, and (b) resources involved with the recovery process for damaged systems, such as time, cost, crews, and material required. System interactions illustrate connections between components categorized into (a) in-sector interactions, which represent network and directivity of connections in a sector, (b) cross-sector interactions, referring to interdependencies between two different sectors, and (c) system state variables that refer to the state of components and entities such as the functionality of an electric substation, recovery timeframe of a component, or the available budget.

*C. Computation Core*

To perform desired tasks and produce outputs of modeled systems, computational algorithms are required to be implemented in simulation modeling. Computation core consists of processes as "mechanisms by which the system and its components make the transition from one state to another over time. Processes dictate how the values of the involved components' state variables change over time" [O'Sullivan-page 5]. As discussed earlier, simulation modeling has the capability to simulate processes in critical infrastructure disaster recovery and estimate timeframes of variable changes. Computation core contains numerical methods to determine the required time, budget, and resources for recovery of a damaged component or sector. It also includes optimization algorithms to support optimal values such as minimum time and resources and optimal number of crews required for recovery. Cost-benefit analysis entails implementation of computational algorithms in this regard. The critical path for recovery of targeted components or clients can be determined by implementation of appropriate shortest path methods depending on the type of directivity of connections. Similarly, evaluation of system resilience from a redundancy perspective entails employing corresponding computational methods. Finally, another aspect of computation core that has been frequently mentioned by potential end-users as a necessity is verification of results of human-centered simulation models by simulation of a previous real-world disaster recovery experience.

V. CONCLUSION

Disaster recovery planning for critical infrastructure is complex and heavily reliant on expert judgement. In this paper, we presented its characteristics based on a spectrum of domain- and user-centered dimensions. We discussed the capability of human-centered simulation modeling to simplify the recovery planning process for decision-makers. We proposed a conceptual design framework for design and development of human-centered simulation modeling. This framework consists of three constructs. User interaction represents design features for end-users to interact with simulation modeling. It enables end-users to assign desired odel parameters of hazard status, system status, and resources in models. System representation indicates components, interactions, and state variables of modeled systems. Lastly, computation core contains computational algorithms to perform processes and analyses, and produce desired outputs. This framework helps human-centered simulation modeling developers be informed about the components required to be incorporated in the design and development of models to support end-users. It is worth noting that the use of the framework is focused on planning for recovery of damaged components of communities. However, recovery planning comprises other various aspects that do not fit in this framework such as damage assessment, inter-organizational decision-making hierarchy, public awareness and engagement, and so on. Future studies may explore other aspects of recovery planning and the potential for creating computer-based tools to facilitate those aspects.

ACKNOWLEDGMENT

This work is funded by National Science Foundation's award #1541025.